\documentclass[prl,aps,showpacs,preprint,superscriptaddress]{revtex4}
\usepackage{graphics}
\usepackage{epsfig}
\usepackage{amsmath}
\usepackage{amsfonts}
\usepackage{amssymb}
\usepackage{graphicx}
\usepackage{color}
\usepackage{tabularx}
\usepackage{txfonts}

\begin{document}

\title{Electronic and Magnetic Properties of single Fe atoms on a CuN Surface; Effects of Electron Correlations}

\author{S. K. Panda}
\affiliation{Department of Physics and Astronomy, Uppsala University, Box 516, SE-751 20 Uppsala, Sweden}

\author{I. Di Marco}
\affiliation{Department of Physics and Astronomy, Uppsala University, Box 516, SE-751 20 Uppsala, Sweden}

\author{O. Gr\aa n\"as}
\affiliation{Department of Physics and Astronomy, Uppsala University, Box 516, SE-751 20 Uppsala, Sweden}
\affiliation{School of Engineering and Applied Sciences, Harvard, 29 Oxford Street, Cambridge, MA 02138, U.S.A.}

\author{O. Eriksson}
\email{olle.eriksson@physics.uu.se}
\affiliation{Department of Physics and Astronomy, Uppsala University, Box 516, SE-751 20 Uppsala, Sweden}

\author{J. Fransson}
\affiliation{Department of Physics and Astronomy, Uppsala University, Box 516, SE-751 20 Uppsala, Sweden}

\begin{abstract}
The electronic structure and magnetic properties of a single Fe adatom on a CuN surface have been studied using density functional theory in the local spin density approximation (LSDA), the LSDA+U approach and the local density approximation plus dynamical mean-field theory (LDA+DMFT). The impurity problem in LDA+DMFT is solved through exact diagonalization and in the Hubbard-I approximation. The comparison of the one-particle spectral functions obtained from LSDA, LSDA+U and LDA+DMFT show the importance of dynamical correlations for the electronic structure of this system. Most importantly, we focused on the magnetic anisotropy and found that neither LSDA, nor LSDA+U can explain the measured, high values of the axial and transverse anisotropy parameters. Instead, the spin excitation energies obtained from our LDA+DMFT approach with exact diagonalization agree significantly better with experimental data. This affirms the importance of treating fluctuating magnetic moments through a realistic many-body treatment when describing this class of nano-magnetic systems. Moreover, it facilitates insight to the role of the hybridization with surrounding orbitals. 
\end{abstract}
\pacs{75.75.-c, 75.30.Gw, 71.27.+a}
\maketitle


Magnetic anisotropies are fundamental to understand the nature of magnetic materials, nano-devices and magnetic structures that approach the single atom limit. As they are equally important for simple collinear and non-collinear magnetic structures in stabilizing the ground state properties, anisotropic magnetic parameters have also been shown to be crucial for dynamical control of non-equilibrium quantities, e.g., magnetic resonances, switching phenomena, damping effects and transport properties. Upon approaching the quantum limit, a full comprehensive and predictive theoretical framework for magnetism necessarily includes a quantum mechanical description of the local atomic environment.
\par
In recent years much effort has been put into experimental studies of  atomic scale anisotropies using local probing, e.g., scanning tunneling microscopy/spectroscopy (STM/STS)~\cite{Hirjibehedin2007,PhysRevLett.102.167203,Oberg2014,Rau2014,PhysRevLett.114.106807}, magnetic force microscopy~\cite{Nocera2012}, and mechanically controlled break junctions, as well as with averaging spectroscopy, e.g., angular/spin resolved photoemission spectroscopy~\cite{PhysRevB.63.195407,Sawada2003,Supruangnet2011}, X-ray magnetic circular dichroism (XMCD)~\cite{PhysRevLett.99.177207,PhysRevLett.108.256811,PhysRevB.89.104424,Ye2013}, and X-ray absorption fine structure (XAFS)~\cite{PhysRevB.92.045423}. Theoretically, progress has been made using model Hamiltonians~\cite{NanoLett.9.2414,PhysRevLett.102.256802,PhysRevB.81.115454,PhysRevLett.104.026601,PhysRevLett.111.127203,PhysRevLett.113.237201,PhysRevB.91.205438}, density functional theory (DFT)~\cite{PhysRevB.79.172409,PhysRevB.82.125438,PhysRevB.88.064411}, and more recently also 
by merging these two strategies~\cite{PhysRevB.88.085112}. It is well known that model Hamiltonians can provide effective phenomenological theoretical descriptions whereas DFT is capable to reproduce ground state properties, and also have an element of being material specific, in principle without any experimental data as an input.
Calculations based on single reference DFT typically fails for materials with strong correlation, something that is within the capabilities of DMFT coupled to accurate electronic structure methods~\cite{RevModPhys.78.865}. Such LDA+DMFT approach has been very successful in addressing strongly correlated electrons systems. However, typically these calculations are aimed at describing bulk properties with good accuracy~\cite{RevModPhys.78.865}, while the properties of isolated paramagnetic defects thus far has remained beyond reach. Hence most of the theoretical analysis of isolated ad-atoms on surfaces has been based on DFT using simple parametrizations of the exchange correlation functional, such as the local spin-density approximation (LSDA) and the generalized gradient approximation (GGA). There are some exceptions to this trend~\cite{PhysRevB.88.085112,PhysRevB.84.115435,Ferron2015}. For example, in the recent work of Mazurenko \textit{et al}~\cite{PhysRevB.88.085112}  spectral properties and exchange interactions of transition metal ad-atoms dimers deposited on the CuN surface were calculated through an Anderson impurity model, where just like LDA+DMFT dynamic correlation effects are considered.
\par
So far there has been a large body of experimental investigations of magnetic nano-structures and ad-atoms on substrates. These studies involve e.g. Co atoms on a Pt substrate~\cite{PhysRevLett.114.106807}, molecular magnets on a transition metal substrate~\cite{PhysRevB.85.155437}, complex chiral magnetic structures of surfaces~\cite{Bode2007,PhysRevLett.110.177204} and quantum corrals~\cite{Heller2008}. All these investigations have been analyzed theoretically, albeit only on an LSDA/GGA or LSDA+U level~\cite{Bonski2009,PhysRevB.85.155437,PhysRevLett.94.187201}, which do not offer a dynamical treatment of strong correlation effects. One may however suspect that a theoretical treatment that goes beyond LSDA/GGA or LSDA+U would bring forth important effects, that could explain e.g. the too small orbital moment of Co atoms on Pt~\cite{Bonski2009}, or the difficulty in describing the magnetic excitation spectrum of Fe on CuN~\cite{PhysRevB.79.172409}. The present work is focused on Fe adsorbed on CuN, as an archetype in this class of nano-magnetic systems. We address details of the electronic structure in relation to spectroscopic data and magnetic properties and while we draw conclusions specific to this system in light of Ref. \onlinecite{Hirjibehedin2007}, we analyze the implications of our results in more general terms. In this Letter we study a single paramagnetic adatom (Fe) on a surface (CuN), using a newly developed scheme based on DMFT combined with a full-potential linear muffin-tin orbital method (FP-LMTO)~\cite{FPLMTO_Orig,FPLMTO}, and calculate all parameters pertaining to a quantum spin Hamiltonian and associated excitation spectrum. Our results show that correlation effects are in general important for this class of nano-magnets, both for electronic structure and magnetic properties.
\begin{figure}[t]
\includegraphics[width=0.99\columnwidth]{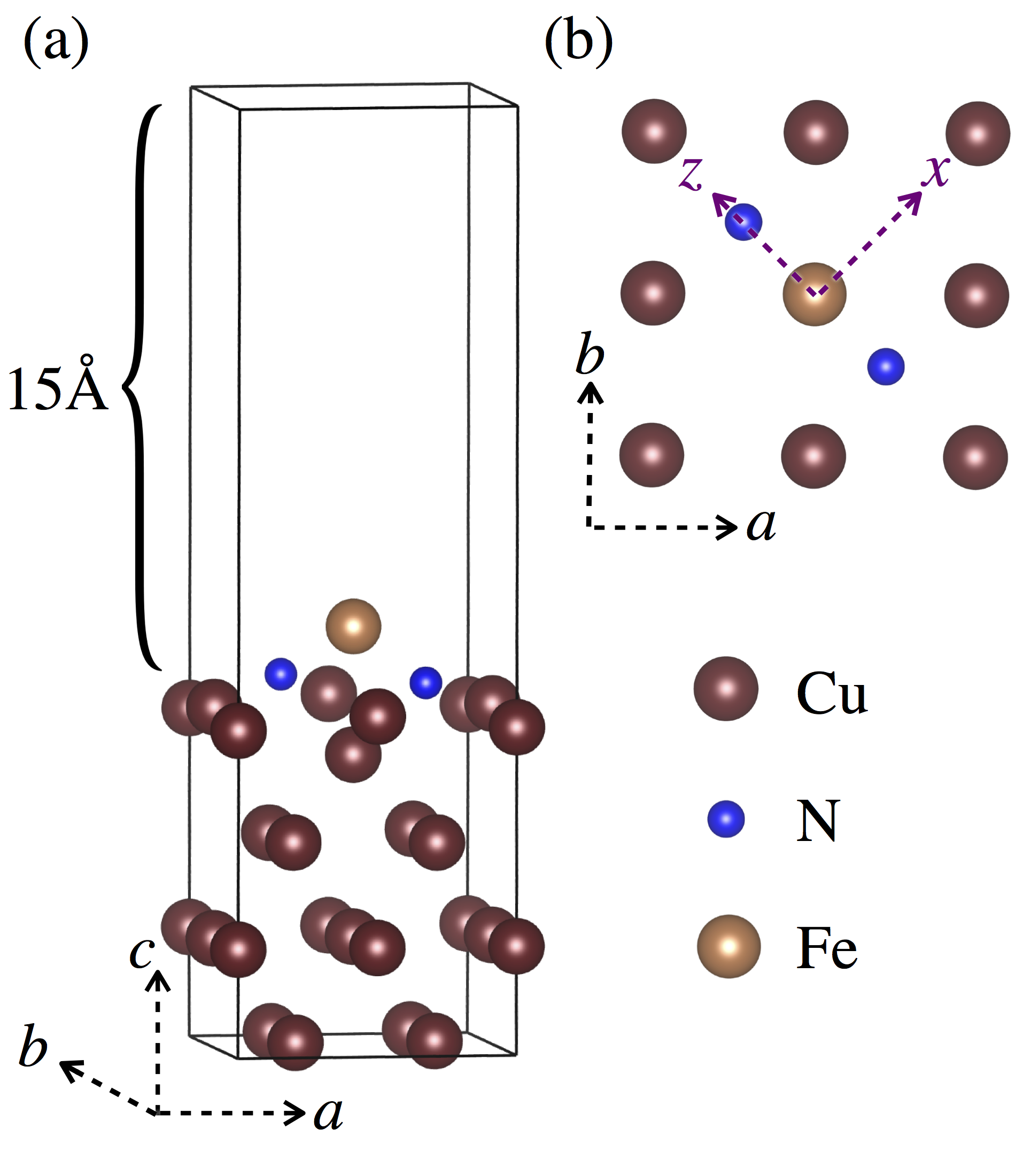}
\caption{(a) Crystal structure of Fe atom on on Cu(100)-c(2$\times$ 2)N surface. (b) A top view of the structure, showing the local axes.}
\label{struct}
\end{figure}
\par 
The experimentally reported structure of an Fe atom on Cu(100)-c(2$\times$ 2)N surface has been simulated using a symmetric slab model, considering a 2$\times$2 supercell of 4 Cu(100) layers and a 15~\r{A} vacuum region, similar to previous DFT study~\cite{PhysRevB.79.172409} (see Fig.~\ref{struct}(a)). The N ions are distributed uniformly on the topmost layer with a 2:1 ratio and an Fe atom is adsorbed to the Cu site as reported in the STM study of Ref.~\onlinecite{Hirjibehedin2007}. We started our investigation by relaxing the ionic positions, using the projector augmented wave (PAW) method~\cite{PhysRevB.50.17953} as implemented in the Vienna \textit{ab-initio} simulation package (VASP)~\cite{PhysRevB.47.558,PhysRevB.54.11169}. The relaxed geometry obtained in our calculation show that the adsorption of the Fe atom induces a local distortion on the surface, in good agreement with an earlier report~\cite{PhysRevB.79.172409}. The N atoms move upwards by 0.52~\r{A} from the topmost Cu-plane and the Cu atom just below the Fe ion is pushed downwards by 0.87~\r{A}, making the Fe-Cu vertical distance 2.32~\r{A}. The vertical height of Fe adatom with respect to the N and topmost Cu plane become 0.93~\r{A} and 1.45~\r{A} respectively. The top view of the structure is shown in Fig.~\ref{struct}(b) where the local $x$ and $z$ axes are respectively set along the (1 1 0) and (-1 1 0) directions. The $y$ axis is chosen along the out-of-plane direction (0 0 1). This local frame is the same as the one used in Ref.~\onlinecite{Hirjibehedin2007}, which will facilitate the comparison between experimental data and our results on the magnetic anisotropy.

The optimized structure has been used to analyze the electronic structure and the magnetic properties within LDA/LSDA, LSDA+U and LDA+DMFT approaches using the FP-LMTO method~\cite{FPLMTO_Orig,FPLMTO} as implemented in the RSPt code~\cite{FPLMTOCode}. The LDA+DMFT calculations of the paramagnetic phase are based on the implementation presented in Refs.~\onlinecite{DMFT_Rspt1,DMFT_Rspt2,DMFT_Rspt4}. As we shall see below, this allows an accurate determination of parameters describing the magnetic anisotropy of an effective quantum spin-Hamiltonian of this class of systems. The effective impurity problem for the Fe-3$d$ states is solved through the exact diagonalization (ED) method~\cite{Marco2012} and also within the Hubbard I approximation (HIA)~\cite{PhysRevB.57.6884,DMFT_Rspt3}. To describe the electron-electron correlation, we have assumed {$U$ = 6 eV} and {$J$ =1.0 eV} for the Fe-$d$ states, in agreement with a previous model study~\cite{PhysRevB.88.085112}. Further technical details has been described in the supplementary material (SM).
\par 
\begin{figure}[t]
\includegraphics[width=0.99\columnwidth]{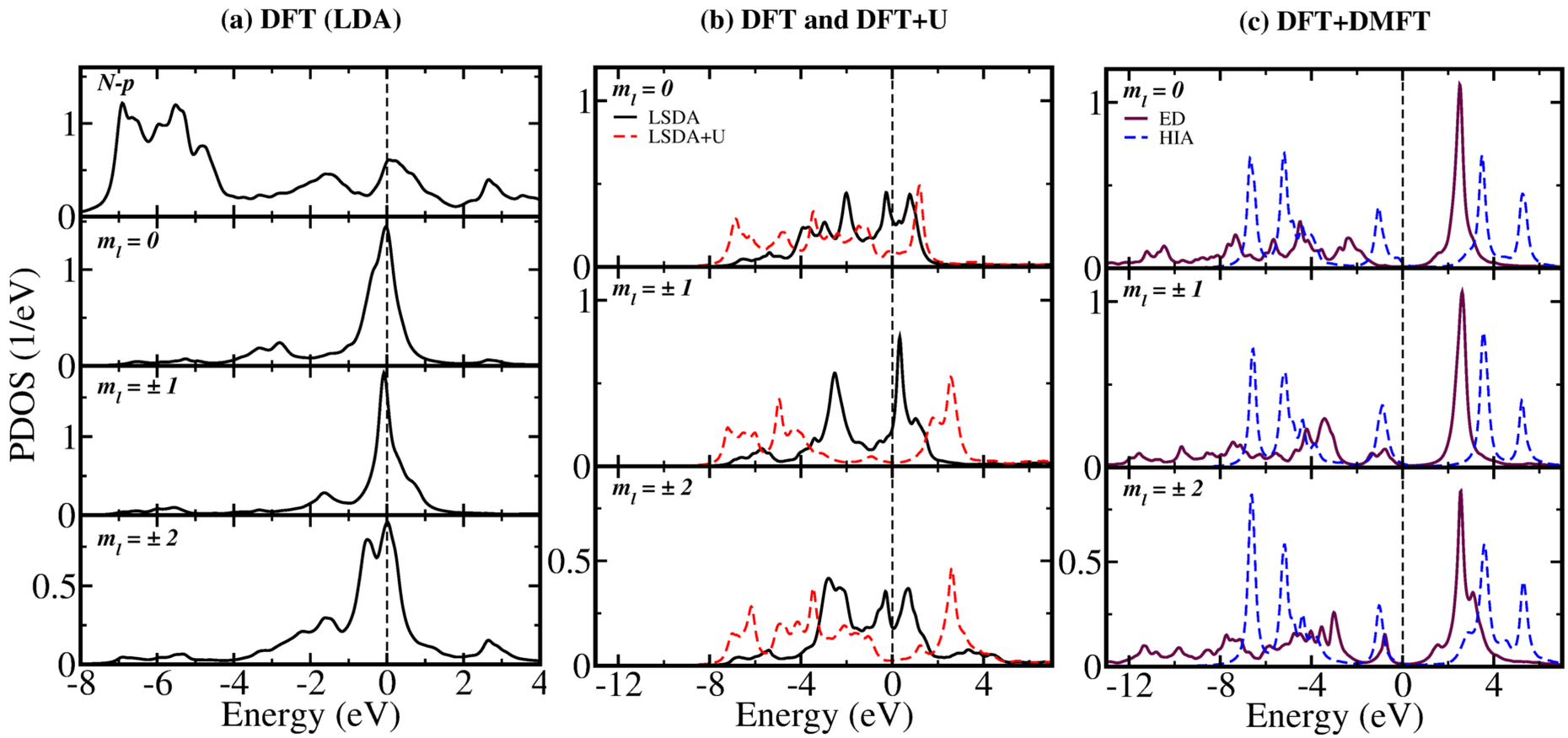}
\caption{PDOS for the Fe-$3d$ orbitals as obtained from LDA (panel a), LSDA and LSDA+U (panel b), LDA+DMFT with HIA and ED (panel c). For LDA also the N-$p$ states are reported.}
\label{PDOS}
\end{figure}
\par 
We first look at the projected density of states (PDOS) obtained using LDA method, as shown in Fig. \ref{PDOS}(a). The Fe-$3d$ states are strongly $m_l$ dependent and especially the curves for $m_l=\pm2$ deviate from the corresponding plots for $m_l=\pm1$ and $m_l=0$ states. Despite these variations, all states have a peak near the Fermi level which are much narrower than compared to bulk bcc Fe, suggesting that the Fe-$3d$ electrons are strongly localized which make LDA/LSDA based approach inappropriate. The peak at the Fermi level emerges from hybridization between the Fe adatom and the N-$p$ states (Fig. \ref{PDOS}(a)) and the hybridization strength is also different for different $m_l$ derived state as confirmed by the computed hybridization function, shown in SM.

\par 
A proper description of the fluctuating moment is outside of the capabilities of Kohn-Sham DFT. Both LSDA and LSDA+U works with symmetry broken solutions, hence a small but finite spurious static magnetic moment is induced in the neighboring atoms. With this in mind we inspect the PDOS obtained with LSDA and LSDA+U, reported in Fig.~\ref{PDOS}(b). Interestingly, the PDOS obtained in LSDA does not exhibit any gap at the Fermi level. Including a static solution to the impurity Hamiltonian within LSDA+U, the one-particle excitation spectrum becomes wider and the spectral weight at the Fermi level decreases. However, the spectrum is still gapless. In Fig.~\ref{PDOS}(c), we report the spectral function obtained in LDA+DMFT, which for simplicity we will also label as PDOS. We note that our paramagnetic LDA+DMFT results closely mimic the experimental scenario, since the system has only a single magnetic impurity with fluctuating local moments and no local Weiss field present. If no hybridization is considered, as in the HIA~\cite{DMFT_Rspt3}, a large gap arises and sharp peaks are present. As expected, the differences between the various $m_l$ states are minor, due to that they originated mainly from the hybridization with the substrate. These effects are taken into account in LDA+DMFT with ED. In Fig.~\ref{PDOS}(c) one can see that the hybridization affects the different $m_l$ projections to a different extent. In particular, in ED the band gap is decreased with respect to HIA, implying that the hybridization with the surface states shifts conduction and valence levels towards the Fermi level and the gap is different for different $m_l$ states. These differences are reflected in the strong magnetic anisotropy, as discussed below. Notice that the low-energy states defining the gap resemble those obtained in LDA+U, which points to a good description of the hybridization function~\cite{Marco2012}. Interestingly, the formation of high-energy satellites in valence band spectra makes the ED spectrum much wider compared to all the other methods. The large differences obtained in LDA+DMFT with ED with respect to the other methods underlines the need of a proper treatment of correlation effects, hybridization and magnetic order to address this class of nano-magnets.   

\bgroup
\def\arraystretch{1.2}
\begin{table}[b]
\caption{Energy for a given magnetization direction with respect to magnetization along $z$ ($\Delta_E$), as well as spin moment $\mu_s$, orbital moment $\mu_o$ and total cell moment $\mu_{tot}$.}
\vspace{0.1cm}
\begin{tabular}{| c | c c c | c c c | }
\hline
 & \multicolumn{3}{c|}{LSDA+SOC} & \multicolumn{3}{c|}{LSDA+U+SOC} \\
  & \hspace{0.03cm} x axis \hspace{0.03cm} & \hspace{0.03cm} y axis \hspace{0.03cm} & \hspace{0.03cm}  z axis \hspace{0.03cm} & \hspace{0.03cm} x axis \hspace{0.03cm} & \hspace{0.03cm} y axis \hspace{0.03cm} & \hspace{0.03cm}  z axis \hspace{0.03cm} \\
\hline
$\Delta_E$ (meV/cell)  &  1.64 & 0.92 & 0.00 & 2.17 & 0.89 & 0.00 \\
$\mu_s$ ($\mu_B$/Fe) & 2.71 & 2.71 & 2.71 & 3.12 & 3.12 & 3.12  \\
$\mu_o$ ($\mu_B$/Fe)  & 0.02 & 0.06  & 0.14 & 0.03 & 0.11 & 0.22 \\
$\mu_{tot}$ ($\mu_B$/cell) & 3.44 & 3.48 & 3.56 & 3.66 & 3.74 & 3.86 \\
\hline
\end{tabular}
\label{moments}
\end{table}
\egroup

%
%

\bgroup
\def\arraystretch{1.2}
\begin{table}[t]
\caption{Magnetic anisotropy parameters $D$ and $E$, obtained through LSDA+SOC and LSDA+U+SOC, and compared to experimental data~\cite{Hirjibehedin2007} and a previous theoretical study~\cite{PhysRevB.79.172409}.}
\centering \vspace{0.1cm}
\begin{tabular}{|c|c|c|c|c|}
\hline
 & LSDA+SOC & LSDA+SOC & LSDA+U+SOC & Experiment \\
 & (from Ref.~\onlinecite{PhysRevB.79.172409}) & (this work) & (this work) & (from Ref.~\onlinecite{Hirjibehedin2007})  \\
 \hline
 $D$ (meV) & -0.36 & -0.32 & -0.38 & -1.55 \\
$E$ (meV) & \hphantom{-}0.10 & \hphantom{-}0.09 & \hphantom{-}0.16 & \hphantom{-}0.31 \\
\hline
\end{tabular}
\label{anisotropy}
\end{table}
\egroup

\bgroup
\def\arraystretch{1.2}
\begin{table}[b]
\caption{Spin excitation energies in meV, obtained within various approaches and compared to the experimental values from Ref.~\onlinecite{Hirjibehedin2007}.}
\centering \vspace{0.1cm}
\begin{tabular}{|c|c|c|c|c|}
\hline
 & \hspace{0.15cm} E$_1$ \hspace{0.15cm} & \hspace{0.15cm} E$_2$ \hspace{0.15cm} & \hspace{0.15cm} E$_3$ \hspace{0.15cm} & \hspace{0.15cm} E$_4$ \hspace{0.15cm} \\
 \hline
 LSDA+SOC & 0.07 & 0.76 & 1.30 & 1.42 \\
 LSDA+U+SOC & 0.18 & 0.84 & 1.80 & 1.88 \\
LDA+DMFT+SOC (HIA) & 0.03 & 12.08 & 12.23 & 13.76 \\
LDA+DMFT+SOC (ED) & 0.49 & 5.32 & 6.34 & 7.68 \\
 Experiment~\cite{Hirjibehedin2007} & 0.18 & 3.90 & 5.76 & 6.56 \\
\hline
\end{tabular}
\label{SpinExcitation}
\end{table}
\egroup
\par 
Next we analyze the magnetic properties of this system, first as obtained from first-principle simulations and then in terms of an effective, quantum spin-Hamiltonian. The computed energies, the Fe spin and orbital moments, as well as the total moments per unit cell are reported in Table~\ref{moments}, for the three different magnetization directions, as obtained from LSDA and LSDA+U with the inclusion of spin-orbit coupling (+SOC). Our calculations within both approaches suggest that the $z$-axis, (line of N ions), is the easy axis of magnetization which is in agreement with experiment~\cite{Hirjibehedin2007} as well as with an earlier DFT study within the LSDA+SOC approach~\cite{PhysRevB.79.172409}. Our calculations reveal that the spin moment of the Fe ion remains unaltered as we change the magnetization directions. However, the orbital moment undergoes a significant change, and is greatest along the easy axis of magnetization. Such a large orbital moment anisotropy is expected from the analysis of Ref.~\onlinecite{PhysRevB.39.865}. Here a direct proportionality between the orbital moment anisotropy and the magneto crystalline anisotropy was derived, and since the magneto crystalline anisotropy of the presently investigated system is large, the orbital moment anisotropy follows. Table~\ref{moments} also show that both the spin and orbital moments increase upon including static Coulomb corrections, as expected. The magnetic anisotropy parameters in LSDA+SOC and LSDA+U+SOC can be obtained by mapping the total energy differences between different magnetization axes (Table~\ref{moments}) to the following spin Hamiltonian:  
\begin{align}
H = g\mu_B{\bf B}\cdot{\bf S} + DS_z^2 + E(S_x^2 - S_y^2) \: .
\label{eq:spinh1}
\end{align}
Here, the first term corresponds to the Zeeman splitting due to the applied magnetic field ${\bf B}$, while the second and third term correspond to the axial and transverse anisotropy energies. Since Fe is very close to the $d^6$ atomic-like configuration, we can assume $S = 2$ in Eq.~(\ref{eq:spinh1}). The computed parameters ($D$ and $E$) obtained from our calculations as well as previously reported theoretical and experimental values are shown in Table~\ref{anisotropy}. Our results within LSDA+SOC are in good quantitative agreement with those by Shick {\itshape{et al.}}~\cite{PhysRevB.79.172409}. We also find that  static Hartree-Fock corrections as described within LSDA+U  do not affect the magnetic anisotropy significantly and the obtained values are far from the experiment, which shows unequivocally that LSDA and LSDA+U are not suitable for describing this class of compounds, even if an artificial magnetic order is assumed.

\par
\begin{figure}[t]
\includegraphics[width=0.99\columnwidth]{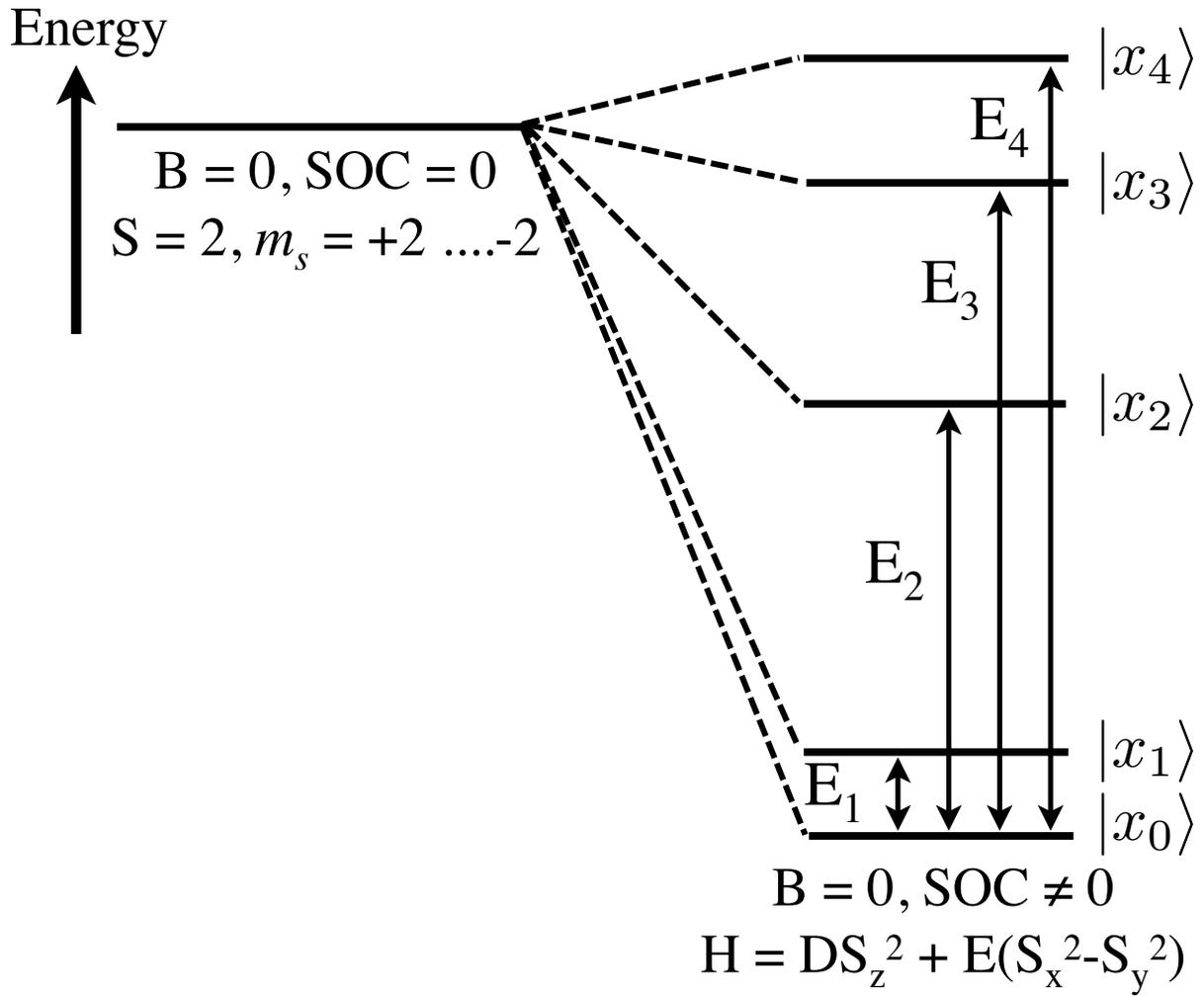}
\caption{Level diagram illustrating how a $S = 2$ degenerate ground state is splitted into five spin states due to spin-orbit coupling at zero magnetic field.}
\label{LevelDiagram}
\end{figure}

Finally we discuss the most important aspect of our study, the estimation of the spin excitation energies via LDA+DMFT. Instead of calculating $D$ and $E$, which are obtained by fitting the measured magnetic field dependent spin excitation energies, we focus on the spin excitation energies for zero magnetic field (${\bf B}$ in Eq.~(\ref{eq:spinh1})), which are measured directly from an STM experiment. 
In absence of any external magnetic field (${\bf B} = 0$), the axial term of the spin Hamiltonian of Eq.~(\ref{eq:spinh1}) will split the degeneracy of the $m_s$ projected states and the transverse term will mix them. Thus the degenerate $S =2$ ground state will be split into five eigenstates, as schematically explained in Fig.~\ref{LevelDiagram}. For the experimentally reported values of $D$ (-1.55 meV) and $E$ (0.31 meV) from Ref.~\onlinecite{Hirjibehedin2007}, these five eigenstates have energies of -6.38, -6.20, -2.48, -0.62 and 0.18 meV respectively (see SM for detail). 

In addition, we estimated the spin excitation energies for the zero magnetic field by diagonalizing the Hamiltonian in Eq.~(\ref{eq:spinh1}) for the values of $D$ and $E$ that we obtained from our LSDA+SOC and LSDA+U+SOC calculation (see Table~\ref{anisotropy}). The excited states can again be marked as in Fig.~\ref{LevelDiagram}. The energy differences with respect to the spin ground state ($|x_0\rangle$) are displayed in Table~\ref{SpinExcitation}. The values obtained with LSDA+SOC strongly underestimate the experimental values reported in Ref.~\onlinecite{Hirjibehedin2007}. The electronic localization associated to the static Coulomb correction in LSDA+U+SOC leads to an increase of the spin excitation energies, but one can still find a substantial disagreement with experimental values. The tendency to underestimate the experimental response is likely to originate from the artificial magnetic order that one has to assume in these approaches to describe the formation of fluctuating local moments. Imposing long-range order, albeit in a different spin state, requires much less energy than creating decoupled local spin excitation. From a methodological point of view the experimental scenario should be much more accessible via LDA+DMFT+SOC. When using ED and HIA, it is possible to extract the spin excitation energies directly from our paramagnetic calculations. The many-body eigenstate $\mathbf{\lvert\psi^{s}_{i,m_s}\rangle}$ arising from the effective impurity can be indexed by $s(s+1) = \langle \mathbf{S^2}\rangle$ and $m_s = \langle S_z\rangle$. The eigenstates corresponding to $m_s = -s, \dots, +s$ are degenerate in energy in absence of SOC. The latter lifts the degeneracy and leads to a set of five eigenstates, which can be directly compared with the spin excitation energies. The values reported in Table~\ref{SpinExcitation} show that the energies obtained in LDA+DMFT+SOC in HIA are much larger than those obtained in LSDA+U+SOC, but also larger than the experimental values. In the Slater Koster model used in Ref.~\onlinecite{PhysRevB.39.865} it is apparent that the crystal field splitting parameter plays a decisive role for the MAE. The emerging picture from DFT+DMFT is that symmetry breaking effects included in the single particle local Hamiltonian, as described by HIA, leads to a correct easy axis, but incorrect spin-excitation spectra. It is crucial to include bath orbitals explicitly in order to get the correct spin-excitation spectra. The reason is the renormalized ligand field splitting and, perhaps most importantly, the allowance of charge fluctuations. Inspecting the thermally averaged $d$-orbital occupation, 6.06 for HIA and 5.46 from ED, it is clear that charge fluctuations are more prominent in ED, with close to half-integer thermally averaged $d$-occupation. This corroborates results from the parametrized impurity model by Ferr\'on \emph{et al}, where they acquire the correct easy axis only when charge fluctuations are included \cite{Ferron2015}. It is clear from Table~\ref{SpinExcitation} that including the hybridization with the substrate is essential for describing the spin excitation spectrum of single atoms of Fe on CuN. This also suggests that the LDA+DMFT, e.g. with the ED solver, is the method of choice to describe correlation effects for nano-magnets in general. In particular, all systems~\cite{Rau2014,PhysRevLett.113.237201,Zhou2010,Khajetoorians2012,Miyamachi2013} with one or a few atoms on a substrate, for which the DOS forms a narrow resonance of a few eV width, are expected to behave similarly to the presently investigated system. 
\par 
In conclusion, we have shown that magnetic order, dynamical correlation effects and hybridization of the transition metal atom with the surface states are very crucial to understand the electronic structure, magnetic properties and spin excitation spectra of Fe on CuN. Static correlation effects as in the LSDA+U method fail to explain the large MAE observed, whereas the LDA+DMFT method describes magnetism of such materials with much improved accuracy. We argue that this should be a rather general feature for ad-atoms of few-atom clusters supported on a substrate, since  the width of all electronic resonances are narrower than in bulk materials, and hence Coulomb repulsion should be more important.

This work was supported by the Swedish Research Council, the KAW foundation, and eSSENCE. Calculations have been performed at the Swedish national computer centers UPPMAX and NSC. 



\end{document}


\title{Supplementary Material\\ 
Electronic and Magnetic Properties of single Fe Atom on a CuN Surface; Effects of Electron Correlations}

\author{S. K. Panda}
\affiliation{Department of Physics and Astronomy, Uppsala University, Box 516, SE-751 20 Uppsala, Sweden}

\author{I. Di Marco}
\affiliation{Department of Physics and Astronomy, Uppsala University, Box 516, SE-751 20 Uppsala, Sweden}

\author{O. Gr\aa n\"as}
\affiliation{Department of Physics and Astronomy, Uppsala University, Box 516, SE-751 20 Uppsala, Sweden}
\affiliation{School och Engineering and Applied Sciences, Harvard, 29 Oxford Street, Cambridge, MA 02138, U.S.A.}

\author{O. Eriksson}
\email{olle.eriksson@physics.uu.se}
\affiliation{Department of Physics and Astronomy, Uppsala University, Box 516, SE-751 20 Uppsala, Sweden}

\author{J. Fransson}
\affiliation{Department of Physics and Astronomy, Uppsala University, Box 516, SE-751 20 Uppsala, Sweden}

\maketitle
\par
\section{Computational Details}
The system has been simulated using a symmetric slab model, considering four (100) layer of fcc Cu. In view of the fact that the Fe adatom will strongly influence the positions of the ions close to it, we have optimized only the positions of the N atoms, the upper two Cu (100) layers and also the Fe atom using a plane wave basis as implemented in Vienna \textit{ab-initio} simulation package (VASP)~\cite{PhysRevB.47.558,PhysRevB.54.11169} with projector augmented wave potentials~\cite{PhysRevB.50.17953}. The Cu-Cu distances for the other two layers has been kept fixed to the bulk values. For the exchange correlation functional, we used local spin-density approximation (LSDA) with generalized gradient corrections of Perdew-Burke-Ernzerhof \cite{PhysRevLett.77.3865}. The Brillouin-Zone integration have been carried out with a 12$\times$ 12$\times$ 1 k-mesh. The kinetic energy cut off of the plane wave basis was chosen to be 600 eV. 
\par 
The optimized structure has been used to analyze the electronic structure and the magnetic properties of this system within LDA/LSDA, LSDA+U and LDA+DMFT approaches. These subsequent calculations are carried out using full potential linearized muffin-tin orbital method (FPLMTO)~\cite{FPLMTO_Orig,FPLMTO} as implemented in the RSPt code~\cite{FPLMTOCode}. We use a tripple-$\kappa$ basis set for the disperse $s-$ and $p-$orbitals, the $3d-$orbitals in the Cu substrate is assigned a double-$\kappa$ basis,  the $3d$-orbitals on the Fe-site is described in single-$\kappa$. 
The effective impurity problem for the Fe $d$ states is described through exact digonalization (ED) method~\cite{Marco2012} and the Hubbard I approximation (HIA)~\cite{PhysRevB.57.6884,DMFT_Rspt3}. The impurity hamiltonian in the ED method can be written as
\begin{equation}
\mathbf{H}^{ED} = \sum_{ij}(\mathbf{H}_{ij}^{LDA}-\mathbf{H}_{ij}^{DC})\mathbf{c}_{i}^{\dagger}\mathbf{c}_{j} + \frac{1}{2}\sum_{ijkl}(\mathbf{U}_{ijkl}\mathbf{c}_{i}^{\dagger}\mathbf{c}_{j}^{\dagger}\mathbf{c}_{l}\mathbf{c}_{k} + \sum_{im}(\mathbf{V}_{im}\mathbf{c}_{i}^{\dagger}\mathbf{c}_{m} + H.C.) + \sum_{m}\mathbf{E}_{m}\mathbf{c}_{m}^{\dagger}\mathbf{c}_{m}, 
\end{equation}
where where the indices $i$, $j$, $k$, $l$ run over the local correlated orbitals and $m$ runs over the auxiliary bath states. Here \textbf{H}$^{LDA}$ is the local projected LDA Hamiltonian,  and \textbf{H}$^{DC}$ is is the double counting correction, which removes from the Kohn–Sham Hamiltonian those contributions that are associated to the added local Coulomb interaction $U$ in order to avoid counting them twice~\cite{RevModPhys.78.865}. The energies $E_m$ and the hybridization strength $V_{im}$ of the auxilary bath states mimic the hybridization of Fe-$3d$ orbitals with all the other states. The Hamiltonian of the finite system is diagonalized numerically to produce an analytical self-energy. The calculations were carried out in the paramagnetic (PM) phase, using a finite temperature (300\,K). 
The convergence of the self-energy has been achieved with 1500 Matsubara frequencies. For the description of electron-electron correlation, the fully rotationally invariant $U$-matrix has been constructed from the Slater parameters $F^0$, $F^2$ and $F^4$. These have been reduced to only two parameters, the Hubbard $U$ and the Hund exchange $J$, by using fixed atomic ratios~\cite{PhysRevB.48.16929}. We have assumed {$U$ = 6 eV} and {$J$ =1.0 eV} for the Fe-d states, in agreement with a previous study~\cite{PhysRevB.88.085112}. For the double counting correction, we used fully localized limit. The fitting of the hybridization function for the ED simulation was done with two bath sites per each 3$d$ orbital. 
\par 
In the HIA method, the last two terms of the Hamiltonian which describe the hybridization effect of the correlated orbital with the other states are absent, implying that the method is only applicable for those orbitals which has negligible hybridization effects. 
\pagebreak
\section{Hybridization function}
\begin{figure}[h]
\includegraphics[width=0.99\columnwidth]{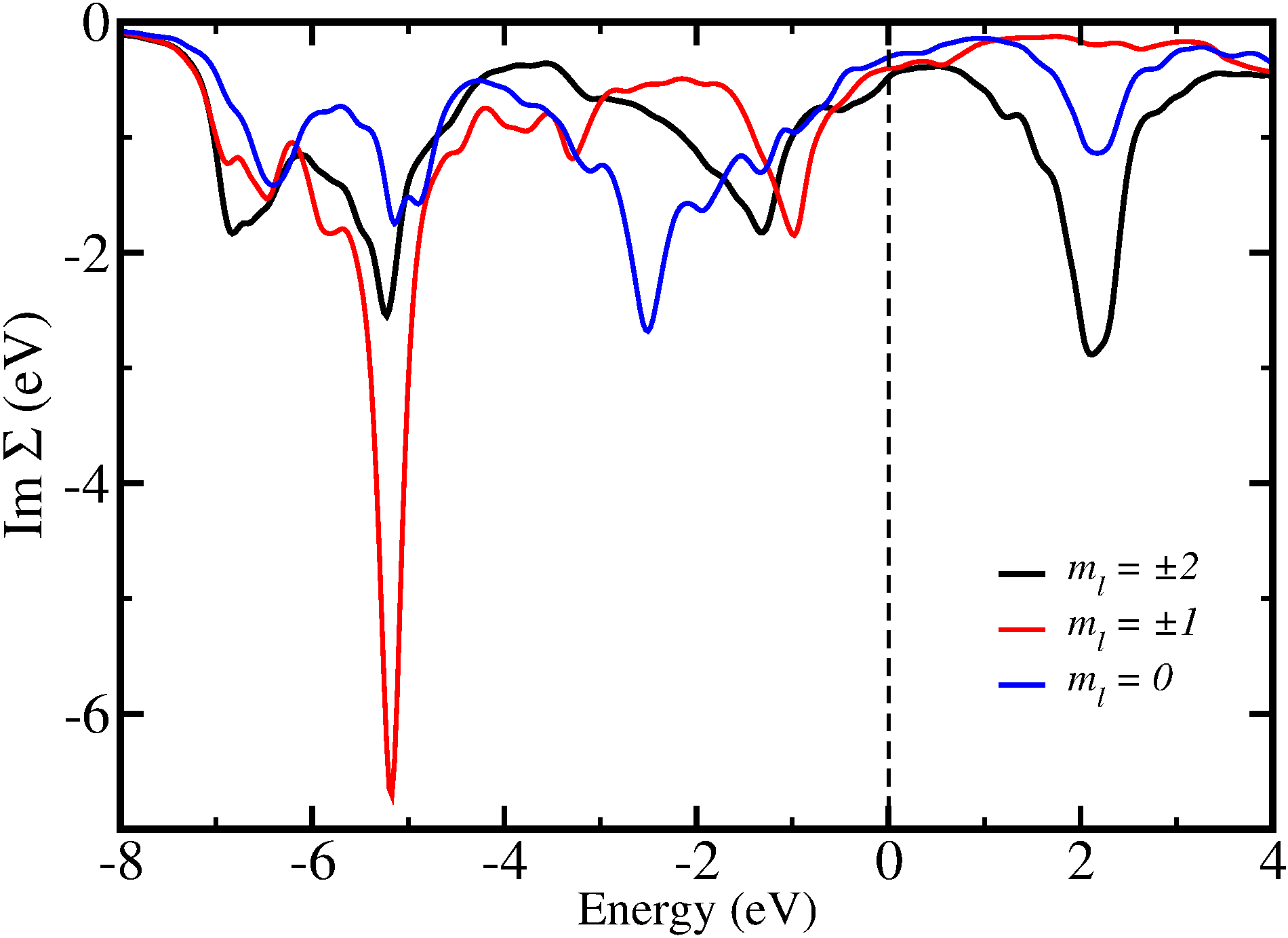}
\caption{Imaginary part of the hybridization function, projected on $m_l$ orbitals.}
\label{hybridization}
\end{figure}

The $m_l$ projected hybridization function is shown in figure~\ref{hybridization}. Our result clearly shows that the hybridization strengths are different for different orbitals and it is the strongest for $m_l = \pm 1$ states. 

\section{Spin Hamiltonian}
To lowest order, spin excitations in an anisotropic environment can be described by the spin Hamiltonian
\begin{align}
H = g\mu_B{\bf B}\cdot{\bf S} + DS_z^2 + E(S_x^2 - S_y^2) \: .
\label{eq:spinh1}
\end{align}
Here, the first term corresponds to the Zeeman splitting due to the applied magnetic field ${\bf B}$, where where g is the $g$-factor and $\mu_B$ is the Bohr magneton. The second and third terms correspond to the phenomenological representations of axial and transverse anisotropy energies, characterized by strengths $D$ and $E$, respectively.  In Equation~\ref{eq:spinh1}, the $z$ axis is chosen along the easy magnetization direction.  The spin Hamiltonian of Eq.~(\ref{eq:spinh1}) can be easily rewritten in terms of the spin raising ($S_+$) and spin lowering operator ($S_-$) in the following form:
\begin{align}
H = g\mu_B{\bf B}\cdot{\bf S} + DS_z^2 + \frac{E}{2}(S_+^2 + S_-^2) \: .
\end{align}
If we again assume Fe to be in a $S = 2$ state, the Hamiltonian in the $m_s$ basis for B = 0 can be written as
\begin{align}
\begin{pmatrix}
4D & 0 & \sqrt{6}E & 0 & 0 \\
0 & D & 0 & 3E & 0 \\
\sqrt{6}E & 0 & 0 & 0 & \sqrt{6}E \\
0 & 3E & 0 & D & 0 \\
0 & 0 & \sqrt{6}E & 0 & 4D
\end{pmatrix}\label{eq:htodiag}
\end{align}
In absence of any external magnetic field (${\bf B} = 0$), the axial term of the above Hamiltonian will split the degeneracy of the $m_s$ projected states and the transverse term will mix them. Thus the degenerate $S =2$ ground state will be split into five eigenstates. For the experimentally reported values of $D$ (-1.55 meV) and $E$ (0.31 meV) from Ref.~\onlinecite{Hirjibehedin2007}, these five eigenstates have energies of -6.38, -6.20, -2.48, -0.62 and 0.18 meV respectively. Their corresponding wavefunctions are
\begin{align}
\begin{array}{rcl}
|x_0\rangle & = & -0.6973|2, +2\rangle + 0.1660|2, 0\rangle - 0.6973|2, -2\rangle\\
|x_1\rangle & = & -0.7071|2, +2\rangle +  0.7071|2, -2\rangle\\
|x_2\rangle & = &  0.7071|2, +1\rangle -  0.7071|2, -1\rangle\\
|x_3\rangle & = & 0.7071|2, +1\rangle +  0.7071|2, -1\rangle\\
|x_4\rangle & = & -0.1174|2, +2\rangle - 0.9861|2, 0\rangle - 0.1174|2, -2\rangle
\end{array}
\end{align}

\bibliography{FeCuNSI}